\documentclass{elsart}
\usepackage{epsfig,latexsym,picinpar,amsmath,amsfonts}

\journal{Information Prcessing Letters}
\volume{69}
\firstpage{283}
\lastpage{289}
\pubyear{1999}

\newcommand{\lastcorrections}%
{{\vskip 0.2in
\begin{sloppypar}
    \baselineskip -0.2in
    \tiny\bf\noindent Last Corrections:\\   
marek: Wed May 27 12:37:31 PDT 1998\\
xtof: Wed May 27 19:09:59 PDT 1998\\
marek: Sat May 30 12:51:24 PDT 1998\\
marek: Sun Jun  7 21:01:02 PDT 1998\\
xtof: Tue Jun  9 19:30:42 PDT 1998\\
xtof: Wed Jun 10 12:36:21 PDT 1998\\
marek: Tue Jun 16 17:50:01 PDT 1998\\
marek: Thu Jun 18 17:31:35 PDT 1998 (cut down to 9 pages)\\
marek: Sat Jun 27 18:01:07 PDT 1998\\
marek: Sun Jun 28 12:26:49 PDT 1998\\
xtof: Mon Jun 29 18:57:22 PDT 1998\\
xtof: Tue Jun 30 12:06:03 PDT 1998\\
xtof: Wed Jul  1 23:35:26 PDT 1998\\
marek: Thu Jul  2 18:58:32 PDT 1998\\
xtof: Fri Jul  3 12:09:56 PDT 1998\\
marek: Fri Jul  3 18:49:24 PDT 1998\\
xtof: Fri Jul  3 22:08:58 PDT 1998\\
marek: Sat Jul  4 13:07:15 PDT 1998\\
xtof: Sun Jul  5 21:16:00 PDT 1998\\
marek: Mon Jul  6 12:05:11 PDT 1998\\
xtof: Mon Jul  6 15:48:10 PDT 1998\\
marek: Tue Jul  7 19:31:19 PDT 1998\\
marek: Wed Jul  8 12:39:11 PDT 1998\\
xtof: Wed Jul  8 15:02:12 PDT 1998 (add Ryser in O(n+m) and condition (*))\\
marek: Wed Jul  8 17:27:33 PDT 1998\\
xtof: Wed Jul  8 18:51:29 PDT 1998\\
xtof: Mon Dec 21 18:58:55 PST 1998 (included some of referee 2's remarks)\\
xtof: Mon Dec 21 19:20:04 PST 1998 (split inequalities r_1<=.. r_k ..>= r_m)
\end{sloppypar}
}}

\newcommand{\rowsum}{\mathit{rowsum}}
\newcommand{\colsum}{\mathit{colsum}}
\newcommand{\hisum}{\mathit{hisum}}
\newcommand{\losum}{\mathit{losum}}

\newcommand{\bfr}{{\bf r}}
\newcommand{\bfc}{{\bf c}}
\newcommand{\bfT}{{\bf P}}

\newcommand{\barT}{{\overline T}}
\newcommand{\barI}{{\overline I}}
\newcommand{\barJ}{{\overline J}}
\newcommand{\barA}{{\overline A}}
\newcommand{\barB}{{\overline B}}
\newcommand{\barC}{{\overline C}}
\newcommand{\barD}{{\overline D}}

\newcommand{\barY}{{\overline Y}}

\newcommand{\braced}[1]{{ \left\{ #1 \right\} }}

\newcommand{\barred}[1]{{ \left| #1 \right| }}

\newcommand{\assign}{{\,\leftarrow\,}}

\newcommand{\suchthat}{{\,:\,}}

\newcommand{\factive}{\alpha}
\newcommand{\lactive}{\beta}

\newcommand{\lext}[2]{{{\mbox{\it lext}}_{#1}(#2)}}
\newcommand{\rext}[2]{{{\mbox{\it rext}}_{#1}(#2)}}
\newcommand{\Ext}[2]{{{\mbox{\it Ext}}_{#1}(#2)}}

\newcommand{\Qed}{\hspace*{1ex}\hfill$\Box$}
\newenvironment{Proof}{{\smallskip\noindent{\bf Proof}\ }}{{\Qed}}

\newenvironment{Algorithm}[1]
	{\ \smallskip\hrule\smallskip\noindent\textbf{Algorithm~{#1}}}
	{\smallskip\hrule\smallskip}

\newcommand{\myparagraph}[1]{{\smallskip\noindent{\bf #1}\ }}

\begin{document}
\begin{frontmatter}
\title{Reconstructing hv-Convex Polyominoes\\ 
	from Orthogonal Projections}

\author[UCR]{Marek Chrobak\thanksref{NSF}}
\author[ICSI]{Christoph D\"urr}

\address[UCR]{Department of Computer Science,	
        University of California,	
\\     	Riverside, CA 92521-0304.	
	\textsl{marek@cs.ucr.edu}}

\address[ICSI]{International Computer Science Institute,	
        	1947 Center Street, Suite 600,			
	        Berkeley, CA 94704-1198.			
	\textsl{cduerr@icsi.berkeley.edu}}

\thanks[NSF]{Research supported by NSF grant CCR-9503498.}


\end{frontmatter}


\section{Introduction}

Tomography is the area of reconstructing objects from projections. In
{\em discrete tomography\/} an object $T$ we wish to reconstruct may be a
set of cells of a multidimensional grid. We perform measurements of
$T$, each one involving a projection that determines the number of
cells in $T$ on all lines parallel to the projection's direction.
Given a finite number of such measurements, we wish to reconstruct $T$
or, if unique reconstruction is not possible, to compute any object
consistent with these projections. Gardner et al.  \cite{GaGrPr97a}
proved that deciding if there is an object consistent with given
measurements is NP-complete, even for three non-parallel projections
in the 2D grid.

In this paper we address the case of orthogonal (horizontal and
vertical) projections of a 2D grid. A given object $T$, defined as a
set of cells in a $m\times n$ grid, can be identified with an $m\times
n$ 0-1 matrix, where the 1's determine the cells of $T$. We will use
all three notations: set-theoretic, integer and boolean, whichever is
most appropriate in a given context.

The {\em row\/} and {\em column sums\/} of an object $T$ are defined
by $\rowsum_i(T) = \sum_j T_{i,j}$ for $i=1,\ldots,m$ and
$\colsum_j(T) = \sum_i T_{i,j}$ for $j=1,\ldots,n$.  The vectors
$\rowsum(T)$ and $\colsum(T)$ represent the horizontal and vertical
projections of $T$. (See Figure~\ref{fig: realiz}.)

\begin{figwindow}[0,r,\epsfig{file=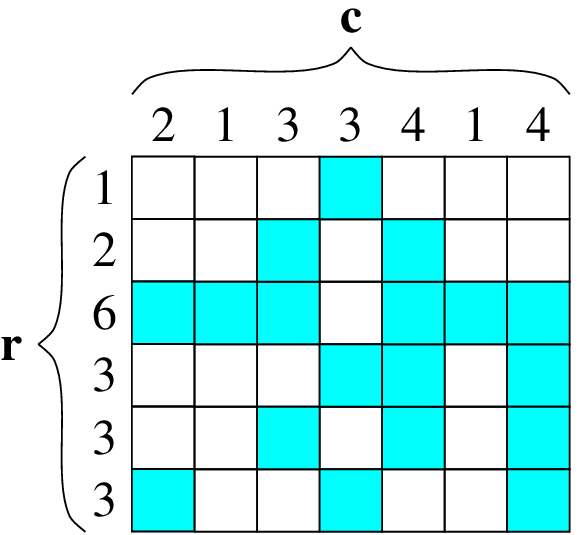,width=4cm},
{An object with its row and column sums.\label{fig: realiz}}] 
Given two integer vectors $\bfr = (r_1,\dots,r_m)$ and $\bfc =
(c_1,\dots,c_n)$, a {\em realization\/} of $(\bfr,\bfc)$ is an object
$T$ whose row and column sum vectors are $\bfr$ and $\bfc$, that is:
$\rowsum(T) = \bfr$ and $\colsum(T) = \bfc$.  In the {\em
reconstruction problem\/}, given $(\bfr,\bfc)$, we wish to find a
realization $T$ of $(\bfr,\bfc)$, or to report failure if such $T$
does not exist.  The corresponding decision problem is called the {\em
consistency problem}.

Properties of 0-1 matrices with given row and column sums have been
studied in the literature since 1950's. We refer the reader to the
work of Ryser \cite{Ryser63}, or to an excellent survey by Brualdi
\cite{Brualdi80}.  Ryser presents an $O(mn)$ algorithm for the
reconstruction problem. In fact, the realizations constructed by his
algorithm can be represented and computed in time $O(m+n)$.

For many objects, the orthogonal projections do not provide sufficient
information for unique reconstruction.  In this case, additional
information about the object's structure could either lead to a unique
realization, or at least substantially reduce the number of
alternative solutions. Some research has been done on {\em
polyominoes\/}, which are connected objects. More formally, we
associate with an object $T$ a graph, whose vertices are the cells of
$T$, and edges join adjacent cells: $((i,j),(i',j'))$ is an edge iff
$|i-i'|+|j-j'|\leq 1$.  If this graph is connected, then $T$ is called
a polyomino.  Woeginger \cite{Woeginger96} proved that the consistency
problem for polyominoes is NP-complete.
\end{figwindow}

Some geometric properties of polyominoes have been studied
too. Call an object $T$ {\em horizontally convex\/} (or {\em
h-convex\/}) if the cells in each row $i$ of $T$ are consecutive, that
is, if $(i,k), (i,l) \in T$ then $(i,j)\in T$ for all $k\leq j
\leq l$. Vertically convex (v-convex) objects are defined analogously.
The consistency problem for h-convex objects (whether we require that
they are polyominoes or not) is also known to be NP-complete
\cite{BaDlNiPi96a}.

If $T$ is both h-convex and v-convex, then we say that $T$ is {\em
hv-convex}. Kuba \cite{Kuba84} initiated the study of hv-convex
polyominoes and proposed a reconstruction algorithm 
with exponential worst-case time complexity.  Quite surprisingly,
as shown later by Barcucci et al \cite{BaDlNiPi96a}, the
reconstruction problem for hv-convex polyominoes can be solved in
polynomial time.  The algorithm given in \cite{BaDlNiPi96a} is,
however, rather slow; its time complexity is $O(m^4n^4)$. In another
paper, Barcucci et al. \cite{BaDlNiPi96b} showed that certain
``median'' cells of $(\bfr,\bfc)$ must belong to any hv-convex polyomino
realization, and, using this result, they developed
an $O(m^2n^2)$-time heuristic algorithm. This new algorithm
is not guaranteed to correctly reconstruct an object, although the
experiments reported in \cite{BaDlNiPi96b} indicate that for randomly chosen
inputs this method is very unlikely to fail.

The main contribution of this paper is an $O(mn\min(m^2,n^2))$-time
algorithm for reconstructing hv-convex polyominoes. Our algorithm is
several orders of magnitude faster than the best previously known
algorithm from \cite{BaDlNiPi96a}, and is also much simpler than the
algorithms from \cite{BaDlNiPi96a,BaDlNiPi96b,Kuba84}. In addition, we
address a special case of {\em centered\/} hv-convex polyominoes, in which
$r_i = n$ for some $i$.  In other words, at least one row is
completely filled with cells.  For this case we show that the
reconstruction problem can be solved in time $O(m+n)$.


\section{hv-Convex Polyominoes}\label{sec: hv-convex polyominoes}

Throughout the paper, without loss of generality, we assume that
$\bfr,\bfc$ denote strictly positive row and column sum vectors.
We also assume
that $\sum_i r_i = \sum_j c_j$, since otherwise $(\bfr,\bfc)$ do not
have a realization. 

An object $A$ is called an {\em upper-left corner region\/} if
$(i+1,j)\in A$ or $(i,j+1)\in A$ implies $(i,j)\in A$.  In an
analogous fashion we can define other corner regions.
By $\barT$ we denote the complement of $T$.  The definition of
hv-convex polyominoes directly implies the following lemma.


\begin{lem}\label{lem: corners}
$T$ is an hv-convex polyomino if and only if $\barT = A\cup B\cup C
\cup D$, where $A$, $B$, $C$, $D$ are disjoint corner regions 
(upper-left, upper-right, lower-left and lower-right, respectively)
such that
{\rm (i)\/} $(i-1,j-1) \in A$ implies $(i,j) \notin D$,
and
{\rm (ii)\/} $(i-1,j+1) \in B$ implies $(i,j) \notin C$.
\end{lem}


Given an hv-convex polyomino $T$ and two row indices $1\le k,l \le m$,
we say that $T$ is {\em anchored\/} at $(k,l)$ if $(k,1), (l,n) \in
T$.  The idea of our algorithm is, given $(\bfr,\bfc)$, to construct a
2SAT expression (a boolean expression in conjunctive normal form with
at most two literals in each clause) $F_{k,l}(\bfr,\bfc)$ with the
property that $F_{k,l}(\bfr,\bfc)$ is satisfiable iff there is an
hv-convex polyomino realization $T$ of $(\bfr,\bfc)$ that is anchored
at $(k,l)$. $F_{k,l}(\bfr,\bfc)$ consists of several sets of clauses,
each set expressing a certain property: ``Corners'' ({Cor}),
``Disjointness'' ({Dis}), ``Connectivity'' ({Con}), ``Anchors''
({Anc}), ``Lower bound on column sums'' ({LBC}) and ``Upper bound on
row sums'' ({UBR}).
	\allowdisplaybreaks
\begin{align*}
\mathit{Cor}&=
\bigwedge_{i,j} \left\{
\begin{array}{*3{rl@{\hspace*{.8em}}}rl}
	A_{i,j} &\Rightarrow A_{i-1,j} &
	B_{i,j} &\Rightarrow B_{i-1,j} &
	C_{i,j} &\Rightarrow C_{i+1,j} &
	D_{i,j} &\Rightarrow D_{i+1,j} \\[-0.2em]
	A_{i,j} &\Rightarrow A_{i,j-1} &
	B_{i,j} &\Rightarrow B_{i,j+1} &
	C_{i,j} &\Rightarrow C_{i,j-1} &
	D_{i,j} &\Rightarrow D_{i,j+1}
\end{array}
\right\}	
		\\
\mathit{Dis}&=
\bigwedge_{i,j}\left\{ X_{i,j}\Rightarrow \barY_{i,j}
\quad\text{: for symbols }X,Y \in\braced{A,B,C,D}, X\neq Y
	\right\}
		\\
\mathit{Con}&=
\bigwedge_{i,j} \left\{
	A_{i,j} \Rightarrow \barD_{i+1,j+1} \hspace*{.8em}
	B_{i,j} \Rightarrow \barC_{i+1,j-1}
	\right\}
		\\[0.2em]
\mathit{Anc}&= 
\barA_{k,1}\wedge \barB_{k,1}\wedge\barC_{k,1} \wedge\barD_{k,1}\:\wedge\:
\barA_{l,n}\wedge \barB_{l,n}\wedge\barC_{l,n} \wedge\barD_{l,n}
		\\
\mathit{LBC}&=
\bigwedge_{i,j} \left\{
\begin{array}{rl@{\hspace*{2em}}rl}
	A_{i,j} &\Rightarrow {\barC_{i+c_j,j}} &
	A_{i,j} &\Rightarrow {\barD_{i+c_j,j}} \\[-0.2em]
	B_{i,j} &\Rightarrow {\barC_{i+c_j,j}} &
	B_{i,j} &\Rightarrow {\barD_{i+c_j,j}}
\end{array}
\right\}
	\quad \wedge\quad 
	\bigwedge_j \left\{ \barC_{c_j,j} \quad \barD_{c_j,j} \right\}
		\\
\mathit{UBR}&=
\bigwedge_j
\left\{
\begin{array}{rl@{\hspace*{1em}}rl}
       \bigwedge_{i \leq \min\braced{k,l}}&
       {\barA_{i,j}}  \Rightarrow B_{i,j+r_i}  &
       \bigwedge_{k \leq i \leq l}&
       {\barC_{i,j}}  \Rightarrow B_{i,j+r_i} \\[-0.2em]
       \bigwedge_{l \leq i \leq k}&
       {\barA_{i,j}}  \Rightarrow D_{i,j+r_i}  &
       \bigwedge_{\max\braced{k,l} \leq i}&
       {\barC_{i,j}}  \Rightarrow D_{i,j+r_i}
\end{array}
\right\}
\end{align*}
Define $F_{k,l}(\bfr,\bfc) = \mathit{Cor}\wedge \mathit{Dis}\wedge
\mathit{Con}\wedge \mathit{Anc} \wedge \mathit{LBC}\wedge
\mathit{UBR}$.  All literals with indices outside the set
$\braced{1,\dots,m}\times\braced{1,\dots,n}$ are assumed to have value 1.


\begin{figure}[ht]
\begin{Algorithm}{1}						\\
{Input:} $\bfr\in\Nset^m$, $\bfc\in\Nset^n$			\\
{W.l.o.g assume:} 
	$\forall i:r_i\in [1,n]$, $\forall j: c_j\in[1,m]$, 
	$\sum_i r_i = \sum_j c_j$ and $m\le n$.
\begin{tabbing}
\textbf{For} \=$k,l = 1,\dots,m$ \textbf{do begin}		\\
\>\textbf{If} $F_{k,l}(\bfr,\bfc)$ is satisfiable, 		\\
\>\textbf{then}	output $T = \overline{A\cup B \cup C \cup D}$
		and \textbf{halt}.				\\
\textbf{end}							\\
output ``failure''.
\end{tabbing}
\end{Algorithm}
\end{figure}


\begin{thm}\label{thm: reduction to 2sat}
$F_{k,l}(\bfr,\bfc)$ is satisfiable if and only if $(\bfr,\bfc)$ have
a realization $T$ that is an hv-convex polyomino anchored at $(k,l)$.
\end{thm}


\begin{Proof}
$(\Leftarrow)$\ If $T$ is an hv-convex polyomino realization of
$(\bfr,\bfc)$ anchored at $(k,l)$, then let $A,B,C,D$ be the corner
regions from Lemma~\ref{lem: corners}.  By Lemma~\ref{lem: corners},
$A$, $B$, $C$ and $D$ satisfy conditions ({Cor}), ({Dis}) and
({Con}).  Condition ({Anc}) is true because $T$ is anchored
at $(k,l)$.  ({LBC}) and ({UBR}) hold because $T$ is a
realization of $(\bfr,\bfc)$.

$(\Rightarrow)$\ If $F_{k,l}(\bfr,\bfc)$ is satisfiable, take $T =
\overline{A\cup B \cup C \cup D}$. Conditions ({Cor}), ({Dis})
and ({Con}) imply that the sets $A$, $B$, $C$, $D$ satisfy
Lemma~\ref{lem: corners}, and thus $T$ is an hv-convex polyomino.
Also, by ({Anc}), $T$ is anchored at $(k,l)$.

It remains to show that $T$ is a realization of $(\bfr,\bfc)$.
Conditions ({UBR}) and ({LBC}) imply that $\rowsum_i(T)\le
r_i$ for each row $i$, and $\colsum_j(T)\ge c_j$ for each column
$j$. Thus \[\sum_i r_i \ge \sum_i\rowsum_i(T) = \sum_j \colsum_j(T) \ge
		\sum_j c_j.\]  
Since $\sum_i r_i = \sum_j c_j$, $T$ must be a
realization of $(\bfr,\bfc)$, and the proof is complete.
\end{Proof}

Each formula $F_{k,l}(\bfr,\bfc)$ 
has size $O(mn)$ and can be computed in time $O(mn)$.
Since 2SAT can be solved in linear time \cite{AsPlTa79,EvItSh76}, we
obtain the following result.

\begin{thm}\label{thm: algorithm hvc}
Algorithm~{\rm 1} solves the reconstruction problem for hv-convex
polyominoes in time $O(mn\min(m^2,n^2))$.
\end{thm}

An implementation (see \cite{CD}) of Algorithm~1 can be made more
efficient by reducing the number of choices for $k,l$.  First,
restrict $k$ to multiples of $c_1$, and $l$ to multiples of
$c_n$. Second, let $m_1$ be the largest index for which $r_1\le \dots
\le r_{m_1}$, and let $m_2$ be the smallest index for which
$r_{m_2}\ge \dots \ge r_m$.  It is easy to see that we can assume that
$\min\braced{k,l} \le m_1$ and $\max\braced{k,l}\ge m_2$ (see
\cite{BaDlNiPi96b}.)  Analogously we define column indices $n_1$ and
$n_2$. With these restrictions on $k,l$, we run the 2SAT algorithm
only $O(\min(m_1m_2/c_1c_n, n_1n_2/r_1r_m))$ times.

One can also remove redundant
clauses from formulas $F_{k,l}(\bfr,\bfc)$. For
example, in the clauses ({Dis}), we only need to state
that regions $A\cap B = \emptyset$ and $C\cap D = \emptyset$. The
other disjointness relations follow from condition ({LBC}). 
Finally, it is not necessary to reconstruct each $F_{k,l}(\bfr,\bfc)$ from
scratch, since only clauses ({UBR}) and ({Anc}) depend on $k$ and $l$.


\section{Centered hv-Convex Polyominoes}\label{sec: centered hv-convex}


\myparagraph{Structure of realizations.}  A \emph{rectangle} $R$ is
the intersection of a set of rows and a set of columns, say $R =
I\times J$, for $I\subseteq\braced{1,\dots,m}$ and $J \subseteq
\braced{1,\dots,n}$.  The rectangle \emph{orthogonal} to $R$ is
$R^\perp = \barI\times \barJ$, where $\barI$ and $\barJ$ are the
complements of $I,J$.

The lemma below follows from the work of Ryser \cite{Ryser63} (see
also \cite{Brualdi80}). We include a simple proof for completeness.


\begin{lem}           \label{lem: separating rectangle}
Let $R = I\times J$ be a rectangle such that
$ \sum_{i\in I}r_i \;=\; \sum_{j\notin J} c_j
			 + \barred{I}\cdot \barred{J}$.
Then each realization $T$ of $(\bfr,\bfc)$ satisfies
$R \subseteq T$ and $R^\perp\cap T = \emptyset$.
\end{lem}

\begin{Proof}
If $T$ is a realization of $(\bfr,\bfc)$ then
\begin{align*}
\barred{R \cap T}  -  \barred{R^\perp \cap T} 
        &= \sum_{i\in I} \sum_{j\in J} T_{i,j}  -
	        \sum_{i\notin I} \sum_{j\notin J} T_{i,j}
\;=\; \sum_{i\in I} \sum_j T_{i,j} - \sum_i \sum_{j\notin J} T_{i,j}\\
&= \sum_{i\in I} r_i - \sum_{j\notin J} c_j 
\;=\; \barred{I} \cdot \barred{J},
\end{align*}
and therefore $R \subseteq T$ and $R^\perp\cap T = \emptyset$.
\end{Proof}


\myparagraph{Representation of hv-convex realizations.} Since we are
interested in \emph{centered hv-convex} realizations, we assume now
that $r_k=n$, for some row $k$. We could also assume that the row sums
satisfy $r_1 \le \dots \le r_k$ and $r_k \ge \dots \ge r_m$,
although our algorithm does not use this assumption.

We represent an hv-convex realization as a vector $T = (t_1,\dots,t_m)$,
where each $t_i\in [1,m-r_i+1]$ is the smallest $j$ for which $(i,j)\in T$.
Thus $(i,j)\in T$ is equivalent to $t_i \le j < t_i+r_i$.  By
definition, such realizations satisfy the horizontal convexity and the
row-sum conditions.  The vertical convexity condition can be written as
\begin{align} 				\tag{VC}\label{VC}
        t_{i+\delta} \le t_i \le t_{i+\delta} + r_{i+\delta} - r_i
\end{align}
where $\delta = 1$ for $i<k$ and $-1$ for $i>k$. The column sums can be
expressed by the $t_i$
as $\colsum_j(T) = |\braced{i \suchthat t_i\le j < t_i+r_i}|$. 


\myparagraph{Partial realizations.}  
Let $X = (t_p,\dots,t_q)$, where $(p,q)\neq (1,m)$ and $k\in[p,q]$,
be a vector in which $t_i\in [1,m-r_i+1]$, for each $i$.
We call column $j$
\emph{unsaturated} if $\colsum_j(X) < c_j$, where $\colsum_j(X) =
|\braced{i \suchthat t_i\le j < t_i+r_i}|$.  Let $\factive_X$ be the
first unsaturated column and $\lactive_X$ the last. (We will omit the
subscript $X$ if $X$ is understood from context.) $X$ is called a
\emph{$[p,q]$-realization} (or a \emph{partial realization}) if it
satisfies condition (\ref{VC}) for $i = p,\dots,q$, and
\begin{align*}
\min\braced{t_p,t_q}\le\factive&\le\lactive
<\max\braced{t_p+r_p,t_q+r_q}, 
						\\
\colsum_j(X) &= c_j\quad \mbox{for}\ j\notin[\factive,\lactive]. 
\end{align*}
If $\max\braced{t_p,t_q} \le \factive$ and
$\lactive<\min\braced{t_p+r_p,t_q+r_q}$, then $X$ is called
\emph{balanced}.  Note that in the definition of partial realizations
we do not insist that
\begin{align}
\colsum_j(X) \le c_j&\quad\mbox{\rm for}\; j = \factive+1,\dots,\lactive-1.
        \label{eqn: pq-realization}
\end{align}
We call $X$ \emph{valid} if it satisfies~(\ref{eqn: pq-realization}).
Clearly, an invalid partial realization cannot be extended to a
realization, but to facilitate a linear-time implementation we will
verify~(\ref{eqn: pq-realization}) only for balanced partial
realizations.

Our algorithm will attempt to construct a realization row by row in the
order of decreasing sums. A $[p,q]$-realization $X$ will be extended
to row $p-1$ or $q+1$, whichever has larger sum.
Each $X$ has at most two extensions, and it may have two extensions
only if it is balanced. The naive approach would be to explore 
recursively all extensions, but, if the number of balanced
realizations is large, it could result in exponential running time.
To reduce the number of partial realizations to explore we show, using
Lemma~\ref{lem: separating rectangle}, that if $(\bfr,\bfc)$ has any
hv-convex realization at all,
then each valid balanced $[p,q]$-realization $X$ can be extended to a
realization of $(\bfr,\bfc)$.  Therefore when we encounter a valid
balanced $[p,q]$-realization, we can discard all other $[p,q]$-realizations.
Consequently, we keep at most two partial realizations at each step.

Suppose that either $q=m$ or $r_{p-1}\ge r_{q+1}$, and
let $\lext{p-1}{X}$, $\rext{p-1}{X}$ be the extensions of $X$
with $t_{p-1}=\factive$ and $t_{p-1}=\lactive-r_{p-1}+1$,
respectively. If $X$ is non-balanced, or balanced and valid,
$\Ext{p-1,q}{X}$ is the set containing these vectors in
$\braced{\lext{p-1}{X},\rext{p-1}{X}}$ which are $[p-1,q$]-realizations.
If $X$ is balanced but not valid then $\Ext{p-1,q}{X} = \emptyset$.
Analogously we define the set
$\Ext{p,q+1}{X}$ for the case when $p=1$ or $r_{q+1}\ge r_{p-1}$.



\begin{Algorithm}{2}
(See Fig.~\ref{fig:algo2})						\\
{Input:} $\bfr\in\Nset^m$, $\bfc\in\Nset^n$				\\
{W.l.o.g. assume:} 
	$\forall i: r_i\in[1,n]$, $\forall j: c_j\in[1,m]$, 
	$\sum_i r_i = \sum_j c_j$ and $\exists k: r_k=n$.
\begin{tabbing}
	$t_k \assign 1$, 
	$X \assign (t_k)$,
	$\bfT \assign \braced{X}$,
	$(p,q) \assign (k,k)$,	 					\\ 
	\textbf{While} \= $(p,q)\neq (1,m)$ \textbf{do begin}		\\
\>	\textbf{If} \= $q=m$ or $r_{p-1}\ge r_{q+1}$ \textbf{then}
		$p\assign p-1$	\textbf{else} $q\assign q+1$	 \\
\>	\textbf{If} $\exists X\in\bfT$ 
		such that $X$ is valid and balanced 			\\
\> 	\textbf{then} $\bfT \assign	\Ext{p,q}{X}$			\\
\>	\textbf{else} $\bfT \assign \bigcup_{X\in\bfT} \Ext{p,q}{X}$	\\ 
	\textbf{end}							\\
	\textbf{If} $\bfT$ contains a realization $T$
		 \textbf{then} output $T$ \textbf{else} output ``failure".
\end{tabbing}
\end{Algorithm}


\begin{figure}[htb]
\centerline{\epsfig{file=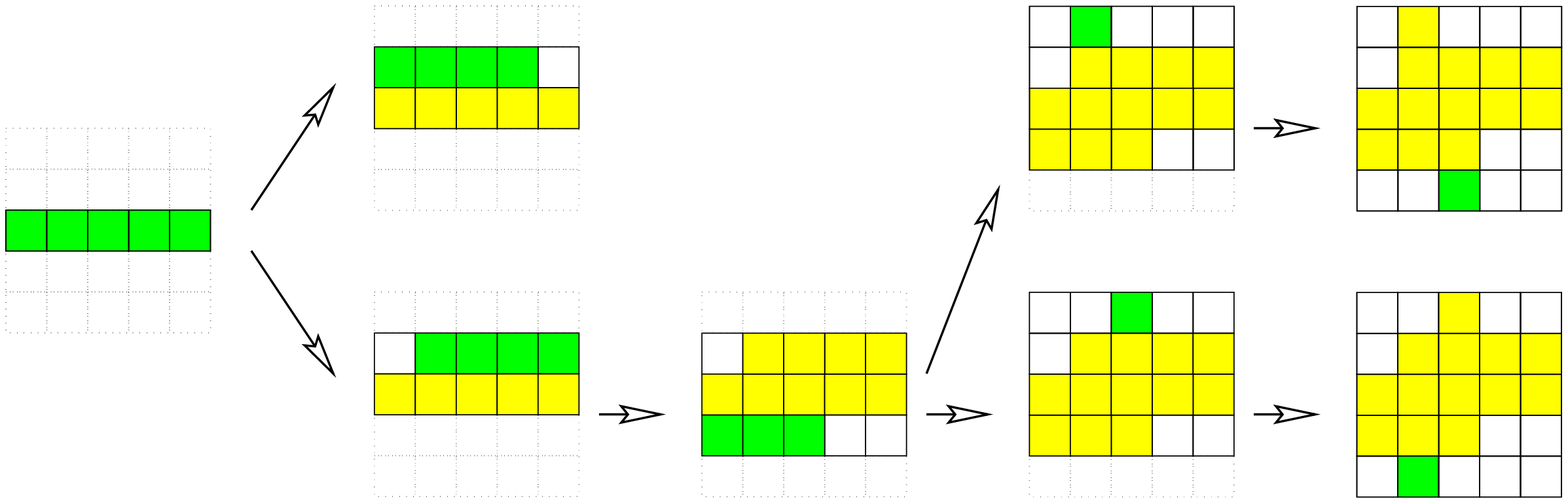,width=13cm}}
\caption{Execution of Algorithm 2 on $\bfr=(1,4,5,3,1)$ and
$\bfc=(2,4,4,2,2)$.}
\label{fig:algo2}
\end{figure}


The correctness of Algorithm~2 follows from the following lemma.


\begin{lem}\label{lem: correctness}
Suppose that $(\bfr,\bfc)$ has an hv-convex realization.  Then, after
each step of Algorithm~{\rm 2}, $\barred{\bfT} \le 2$ and
there is $X\in \bfT$ that can be
extended to an hv-convex realization of $(\bfr,\bfc)$.
\end{lem}

\begin{Proof}
The proof is by induction on $q-p+1$.
Denote by $\bfT_{p,q}$ the set $\bfT$ at the beginning of the
{\bf while} loop. The lemma holds trivially for
$p=q=k$. Assume that it holds for some $\bfT_{p,q}$. 
Without loss of generality, assume that either $q=m$ or
$r_{p-1}\ge r_{q+1}$.

Suppose first that $\bfT_{p,q}$ contains a valid
balanced $[p,q]$-realization $X$.
Let $I = [p,q]$, $J = [\factive_X,\lactive_X]$ and $R=I\times J$.
Then $\sum_{i\in I}r_i = \sum_{j\notin J}c_j + \barred{I}\cdot\barred{J}$.
By Lemma~\ref{lem: separating rectangle}, we get that for each
realization $T = (z_1,\dots,z_m)$,
$R\subseteq T$ and $R^\perp\cap T = \emptyset$.
If we replace $z_p,\dots,z_q$ by
$t_p,\dots,t_q$, we get another realization $T'$ that
is an extension of $X$. Since $q=m$ or
$r_{p-1}\ge r_{q+1}$, $z_{p-1}$ must be either
$\factive_X$ or $\lactive_X-r_{p-1}+1$. Therefore
$T'$ is an extension of one of $\rext{p-1}{X},\lext{p-1}{X}$.
Since $\bfT_{p-1,q} = \Ext{p-1,q}{X}$,
the lemma holds for $\bfT_{p-1,q}$.

If $\bfT_{p,q}$ does not contain a
valid balanced $[p,q]$-realization then, for $X\in\bfT_{p,q}$, either
$\factive_X < \max\braced{t_p,t_q}$ or
$\lactive_X \ge \min\braced{t_p+r_p,t_q+r_q}$.
Thus each $X\in\bfT_{p,q}$ gives rise to at most one
extension in $\bfT_{p-1,q}$, so we have $\barred{\bfT_{p-1,q}}\le 2$.
Pick $X\in\bfT_{p,q}$ that can be extended to a 
realization $T = (t_1,\dots,t_m)$.
We can assume $\factive_X < \max\braced{t_p,t_q}$,
since the other case is symmetric.
If $t_p\le\factive_X<t_q$, then $t_{p-1} = \factive_X$, and $T$ is
an extension of $\lext{p-1}{X}\in \bfT_{p-1,q}$.
If $t_q\le\factive_X<t_p$, then $t_{q+1} = \factive_X$, and thus, by
$t_{p-1}\ge t_{q+1}$, we have $t_{p-1} = \lactive_X-r_{p-1}+1$.
Therefore $T$ is an extension
of $\rext{p-1}{X}\in\bfT_{p-1,q}$.
\end{Proof}


\myparagraph{Linear-time implementation.}
A naive implementation of Algorithm~2 takes time $O(mn)$.  
We show how to reduce the running time to $O(m+n)$.

The computation can be divided into phases, where each phase starts
when a valid balanced partial realization $Y$ is encountered. During
a phase, for each $X\in\bfT$, we store only rows
which are not in $Y$, and when the phase ends the new rows 
are copied into $Y$. We also maintain the
indices $\factive =\factive_X$ and $\lactive = \lactive_X$.
Call a column $j$ \emph{active} if $j\in [\factive,\lactive]$.
To keep track of the sums of the active columns, we use two arrays
$\hisum_j$ and $\losum_j$ which represent, respectively, the column
sums of rows $[p,k-1]$ and $[k+1,q]$ in $X$. 
Only the values of $\hisum_j$ for $j\in[\factive_X,t_p-1] \cup
[t_p+r_p,\lactive_X]$ are stored explicitly. For $j =
[t_p,t_p+r_p-1]$, we know that $\hisum_j = k-p-1$.  When $t_p$
increases or $t_p+r_p$ decreases, we simply fill the new entries with the
correct values. The array $\losum_j$ is maintained analogously.  In this way,
we can maintain these arrays in total time $O(m+n)$ and we can query
$\colsum_j$ in time $O(1)$, since $\colsum_j = \hisum_j + 1 +  \losum_j$.

It remains to show that condition (\ref{eqn: pq-realization}) can be
tested in total time $O(m+n)$. We keep the list $A$ of the active columns
$j$ in the order of increasing sums. Let $\ell$ be the
first column in $A$, that is $c_\ell = \min_{j\in[\factive,\lactive]}
c_j$.  Then for a balanced realization $X$, $(\ref{eqn:
pq-realization})$ holds iff $q-p+1\leq c_\ell$, and thus each
verification of $(\ref{eqn: pq-realization})$ takes time $O(1)$.
Initially, $A$ can be constructed in time $O(m+n)$ by bucket-sort.
Each active column has a pointer to its record in $A$.  When we
increase $\alpha$ or decrease $\beta$, we simply remove from $A$ the columns
that become non-active, at cost $O(1)$ per deletion.


In summary, we obtain the following result.
\begin{thm}
Algorithm~{\rm 2} solves the reconstruction problem for centered hv-convex
polyominoes, and it can be implemented in time $O(m+n)$.
\end{thm}


\section{Final Comments}\label{sec: final comments}

To speed up Algorithm~1 further, one can explore the fact that
the consecutive expressions $F_{k,l}(\bfr,\bfc)$ differ only very slightly.
Thus, an efficient {\em dynamic\/} algorithm for 2SAT could be used to improve
the time complexity. The 2SAT problem is closely related to
strong connectivity in directed graphs, for which no efficient dynamic
algorithms are known.  Nevertheless, it may be possible to
take advantage of the special structure of the 2SAT
instances arising in our algorithm, as well as the fact that the
sequence of clause insertions/deletions in $F_{k,l}(\bfr,\bfc)$ is
known in advance.

It would be interesting to investigate polyominoes which are digital
images of convex shapes. Call a polyomino $T$ \emph{convex} if its
convex hull does not contain any whole cells outside $T$.  How fast
can we reconstruct convex polyominoes?


\bibliographystyle{plain} \bibliography{disctomo}
\end{document}